\documentclass[12pt]{article}
\usepackage{amsmath}
\usepackage{amssymb}
\usepackage[english]{babel}
\usepackage{graphicx}
\usepackage{indentfirst}
\usepackage{mathrsfs}
\usepackage{dsfont}
\usepackage{setspace}
\usepackage[labelfont=bf,labelsep=period]{caption}

\usepackage[numbers,sort&compress]{natbib}

\newtheorem{corollary}{Corollary}
\newtheorem{lemma}{Lemma}

\begin{document}\baselineskip24pt

\title{Nonlocal Correlation of Spin in High Energy Physics}

\author
{Chen Qian$^{1}$, Jun-Li Li$^{2,4}$, Abdul Sattar Khan$^{3}$ and Cong-Feng Qiao$^{3,4\footnote{Corresponding author; qiaocf@ucas.ac.cn}}$ \\ [0.2cm]
\footnotesize{$^{1}$Department of Modern Physics, University of Science and Technology of China, Hefei 230026, China}\\[2pt]
\footnotesize{$^{2}$Center of Materials Science and Optoelectronics Engineering \& CMSOT,} \\
\footnotesize{University of Chinese Academy of Sciences, Beijing 100049, China}\\
\footnotesize{$^{3}$School of Physical Sciences, University of Chinese Academy of Sciences, Beijing 100049, China}\\
\footnotesize{$^{4}$Key Laboratory of Vacuum Physics, University of Chinese Academy of Sciences} \\
\footnotesize{Beijing 100049, China}}

\date{}

\maketitle

\begin{abstract}\doublespacing
Nonlocality is a key feature of quantum theory and is reflected in the violation of Bell inequalities for entangled systems. The experimental tests beyond the electromagnetism and massless quanta are of great importance for understanding the nonlocality in different quantum interactions. In this work, we develop a generalized Clauser-Horne inequality pertaining especially to the high energy physics processes, which is quantum mechanical intervene free. We find, in the process of pseudoscalar quarkonium exclusive decay to entangled $\Lambda\bar{\Lambda}$ pairs, the inequality could be violated and is verifiable in high energy experiments like BES III or BELLE II.
\end{abstract}

\section{Introduction}

Quantum nonlocality, which distinguishes quantum physics from the classical ones, lies at the heart of quantum mechanics. In the studies of quantum theory, the nonlocality is normally tested by the violation of Bell inequalities \cite{Bell,CHSH,CHBI}. Many beautiful experiments have been carried out to this end, most of which rely on the entanglement of photons \cite{Aspect-Exp}. Since it is the fundamental nature of quantum physics, the nonlocality should be examined in various extreme situations rather than limiting it to low energy electromagnetic interaction with massless quanta. Investigation of the nonlocality in high energy physics has attracted more and more attention in recent years \cite{Bertlmann-1, HEPNP-2007}, though the corresponding experiments tend to be tough and delicate schemes are needed to avoid various loopholes.

Testing the nonlocal correlation using spin or polarization in high energy physics has been sorted into four classes based on the interactions in the decay processes \cite{etacVV}. Since the work of Ref.\cite{Tornqvist-1981}, the study on correlation in the baryon decays has been a hot topic in particle theory \cite{Baranov,ptep} and high energy experiments \cite{BES-baryon}. However, so far, controversial issues still remain in testing the nonlocality via high energy processes. First, dichotomic observables are not explicitly available in correlation functions \cite{EPRinHEP}. Second, when embedding the correlation functions into Bell's inequalities, some of the coefficients may experience renormalization in the framework of quantum mechanics, which is somehow self-attesting and improper for the testimony of quantum nonlocality \cite{Hiesmayr-2015}. Third, a conclusive Bell's test requires active control of measurement settings, the so-called free will, but in high energy phenomena, for instance, spontaneous decays, that is to say the ``measurements'' performed, are usually passive \cite{BBGH}.

In this work, we explore the application of the Clauser-Horne (CH) \cite{CHBI} inequality to a high energy experiment, by which the first two problems above can be avoided since here the probabilities rather than correlations are employed. Another superiority, by means of probability density, lies in the ease of its experimental measurement. We will remark on the third problem in the conclusion. In high energy physics experiments, a huge number of quarkonium states are being accumulated at, e.g., BES III, BELLE II or even LHCb detectors. We find the exclusive process of $\eta_c$ to $\Lambda \bar{\Lambda}$ is an ideal process for our aim, where the $\Lambda$ pair is predominantly entangled in spin degrees of freedom with s-wave orbital angular momentum. Noticing the spin measurement in the $\Lambda$ decay amplitude is not simply a probability distribution ranging from $0$ to $1$, we generalize the original CH inequality to a novel form, which is suitable for any kind of decay amplitude.

\section{Generalized CH inequality}

According to the measurement postulate of quantum mechanics, a general quantum measurement is described by a collection $\{M_m\}$ of measurement operators, and the probability for getting the outcome $m$ is given by $p_{m} = \langle \psi|M^{\dag}_m M_m|\psi\rangle$ \cite{QCQI-book}. The probabilities are positive semidefinite and normalized, i.e., $0\leq p_{m} \leq 1$ and $\sum_m E_m = \mathds{1}$ where $E_m \equiv M_m^{\dag}M_m$. Taking the spin 1/2 system as an example, the measurement $\{M_{m}\}$ may be performed by an apparatus along the direction $\vec{n}$, in which each particle triggering the apparatus gives one of the dichotomic results $m=\pm$. As we are only interested in the possibility of each measurement outcome, discussions on the technical details and the configurations of the apparatus are not our main concern. A realistic description of the possibility of the measurement outcome may be formulated as \cite{CHBI}
\begin{align}
P (\vec{n}) = \int_{\Gamma}  p_m(\lambda,\vec{n}) \rho(\lambda) \, \mathrm{d}\lambda\; ,\; a\leq p_m(\lambda,\vec{n}) \leq b \;.\label{general-p-n}
\end{align}
Here, $\lambda$ is the hidden variable determining the possibility $p_m(\lambda,\vec{n})$, by which the particle triggers the apparatus, and $\Gamma$ denotes the space of the hidden variable with a normalized distribution $\rho(\lambda)$; $a$ and $b$ are the lower and upper bounds of the possibility, $0\leq a\leq b\leq 1$, and the subscript of $P_{m}(\vec{n})$ is omitted for convenience. Within a local and realistic theory, we have the following:
\begin{lemma}
In a bipartite system of particles 1 and 2, if we perform the measurement $M_m$ with $p_m(\vec{n}_1) \in [a_1,b_1]$ and $p_m(\vec{n}_2)\in [a_2,b_2]$ on each side, the local realism leads to the following inequality:
\begin{align}
P(\vec{n}_1,\vec{n}_2)- P(\vec{n}_1,\vec{n}'_2)+ P(\vec{n}'_1,\vec{n}_2)+ P(\vec{n}'_1,\vec{n}'_2) & \nonumber \\
-(a_2+b_2)P(\vec{n}'_1)-(a_1+b_1)P(\vec{n}_2)+a_1b_2+ b_1a_2 & \leq 0 \; .\label{GCH-2}
\end{align}
where $P(\vec{n}_1,\vec{n}_2) = \displaystyle \int_{\Gamma} p_m(\lambda,\vec{n}_1) p_m(\lambda,\vec{n}_2) \rho(\lambda)\,\mathrm{d}\lambda$ is the joint distribution and $0\leq a_i\leq b_i\leq 1$. \label{Lemma-1}
\end{lemma}
\noindent Lemma \ref{Lemma-1} can be regarded as a generalized CH inequality for the measurements. The proof of Lemma \ref{Lemma-1} is straightforward by taking advantage of the following inequality \cite{CHBI}:
\begin{align}
x_1y_1-x_1y_2+x_2y_1+x_2y_2 - (a_2+b_2)x_2-(a_1+b_1)y_1+a_1b_2 + b_1a_2 \leq 0 \label{CH-inequ-m}
\end{align}
for parameters $x_{1,2} \in [a_1,b_1]$ and $y_{1,2} \in [a_2,b_2]$.

\section{Violation of new CH inequality in entangled $\Lambda \bar{\Lambda}$}

First we show the local realism predictions for the joint distribution of momenta of $p$ and $\bar{p}$ in the decays of the $\Lambda \bar{\Lambda}$ system. The differential decay width of $\Lambda \to p \pi^-$ with $p$ moving in direction $\vec{n}_p$ is written \cite{Lambda0}
\begin{eqnarray}
\frac{\mathrm{d}\sigma_{\Lambda \to p\pi^-}}{\mathrm{d}\vec{n}_{p}} = \frac{1}{4\pi}(1+\alpha_{-} \vec{s}_{\Lambda} \cdot \vec{n}_{p}) \; . \label{diff-lambda-ppi}
\end{eqnarray}
Here $\vec{s}_{\Lambda} =\mathrm{Tr}[\vec{\sigma}\rho]$ denotes the polarization vector for the spin state $\rho$ of the hyperon $\Lambda$ \cite{Yang-2018}, and $\vec{n}_{p}$ is a unit vector; $\alpha_{-}$ is the decay parameter for $\Lambda$ \cite{pdg}. Without loss of generality, we may choose the polarization vector of $\Lambda$ to be the $z$-axis, then the probability of finding the proton leaving in the direction $\vec{n}_p$ with polar angle $\theta$ is;
\begin{align}
p(\vec{n}_p)  = 2\pi \frac{\mathrm{d}\sigma_{\Lambda \to p\pi^-}}{\mathrm{d}\Omega_{p}} = \frac{1}{2}(1+\alpha_{-} |\vec{s}_{\Lambda}| \cos\theta) \; , \label{p-measurement-model}
\end{align}
which can range from $\frac{1-\alpha_{-}}{2}$ to $\frac{1+\alpha_{-}}{2}$ as the degree of the polarization $|\vec{s}_{\Lambda}|\leq 1$. Analogously, the distribution for the polarized $\bar{\Lambda}$ can also be measured with decay parameter $\alpha_+$. In view of the (approximate) $CP$ conservation arguments, the decay parameters satisfy $\alpha = \alpha_{-}= -\alpha_{+} \simeq 0.750$ \cite{pdg}, hence, $0\leq \frac{1-\alpha}{2} \leq \frac{1+\alpha}{2} \leq 1$. From Lemma \ref{Lemma-1} we have the following:
\begin{corollary}
In a bipartite system consisting of $\Lambda$ and $\bar{\Lambda}$, the local realism predicts that the joint distribution of the momenta of $p$ and $\bar{p}$ satisfies
\begin{align}
P(\vec{n}_1,\vec{n}_2)- P(\vec{n}_1,\vec{n}'_2)+ P(\vec{n}'_1,\vec{n}_2)+ P(\vec{n}'_1,\vec{n}'_2) & \nonumber \\
-P(\vec{n}'_1)-P(\vec{n}_2)+ \frac{1 - \alpha^2}{2} & \leq 0 \; . \label{CH-lambda-bar}
\end{align}
Here $\vec{n}_{1}$, $\vec{n}'_{1}$ are directions of momenta of $p$ and $\vec{n}_{2}$, $\vec{n}'_{2}$ are that of $\bar{p}$; $\alpha$ is the decay parameter of $\Lambda(\bar{\Lambda})$ decay.
\end{corollary}

In quantum theory, the distribution of Eq. (\ref{p-measurement-model}) may be explained by the following measurement process. The Hilbert spaces of the spin of $\Lambda$ and the momentum of protons are coupled by a unitary interaction $U$
\begin{align}
U: |\psi\rangle \otimes |\vec{n}_p\rangle \mapsto M_+(\vec{n}_p)|\psi\rangle \otimes |\vec{n}_p\rangle + M_-(\vec{n}_p)|\psi\rangle \otimes |-\vec{n}_p\rangle \; ,
\end{align}
where $\vec{n}_p$ is the unit vector of the momentum and
\begin{align}
M_{\pm}(\vec{n}_p) \equiv  \displaystyle \frac{1}{\sqrt{2(|S|^2+|P|^2)}} \left[S + P  \vec{\sigma} \cdot (\pm\vec{n}_p) \right] \; . \label{M+lambda}
\end{align}
Here $S$ and $P$ are the decay amplitudes of $\Lambda$ for the $S$ and $P$ waves and $\vec{\sigma}$ represents the spin operator. The parameter $\alpha_-$ relates to the decay amplitudes by the equation $\alpha_{-} = (S^*P+SP^*)/(|S|^2 + |P|^2)$. For $\Lambda$ spinning along the $z$-axis, i.e. $|\psi\rangle = |z\rangle$, the probability for the proton going along $\vec{n}_p$ is
\begin{align}
p(\vec{n}_p) & = \langle z|E_+|z\rangle = \frac{1}{2}(1+\alpha_- \cos\theta)\; , \label{QM-Measurement}
\end{align}
where $E_+ = M_+^{\dag}M_+ = \frac{1}{2}(\mathds{1}+\alpha_-\vec{\sigma} \cdot \vec{n}_p)$ and the argument $\vec{n}_p$ in $M_+$ is suppressed for simplicity. The probability of $p$ coming from the reverse direction of $\vec{n}_p$ is
\begin{align}
p(-\vec{n}_p) = \langle z| E_{-}|z\rangle = \frac{1}{2}(1-\alpha_- \cos\theta) \; .
\end{align}
Here $E_- =M_-^{\dag}M_-$ and $E_+ + E_- =\mathds{1}$. For the $\Lambda \bar{\Lambda}$ system described by a bipartite state $\rho_{12}$, the joint distribution for proton $p$ coming along $\vec{n}_1$ and antiproton $\bar{p}$ coming along $\vec{n}_2$ is
\begin{align}
P(\vec{n}_1,\vec{n}_2) & = \mathrm{Tr}\left[\rho_{12} \left(E_{+}^{(1)} \otimes E_{+}^{(2)} \right)\right] \; , \label{P-n1n2-Q}
\end{align}
where $E_+^{(i)}$ are the measurement operators for particles $\Lambda$ and $\bar{\Lambda}$, and the one-side distribution
\begin{align}
P(\vec{n}_1) = \mathrm{Tr}\left[\rho_{12}\left( E_+^{(1)}\otimes \mathds{1}^{(2)} \right)\right] = P(\vec{n}_1,\vec{n}_2) + P(\vec{n}_1,-\vec{n}_2)\; . \label{P-n1-Q}
\end{align}
Here $\mathds{1}^{(2)} = E_+^{(2)} + E_-^{(2)}$ is employed.

\begin{figure}\centering
\scalebox{0.4}{\includegraphics{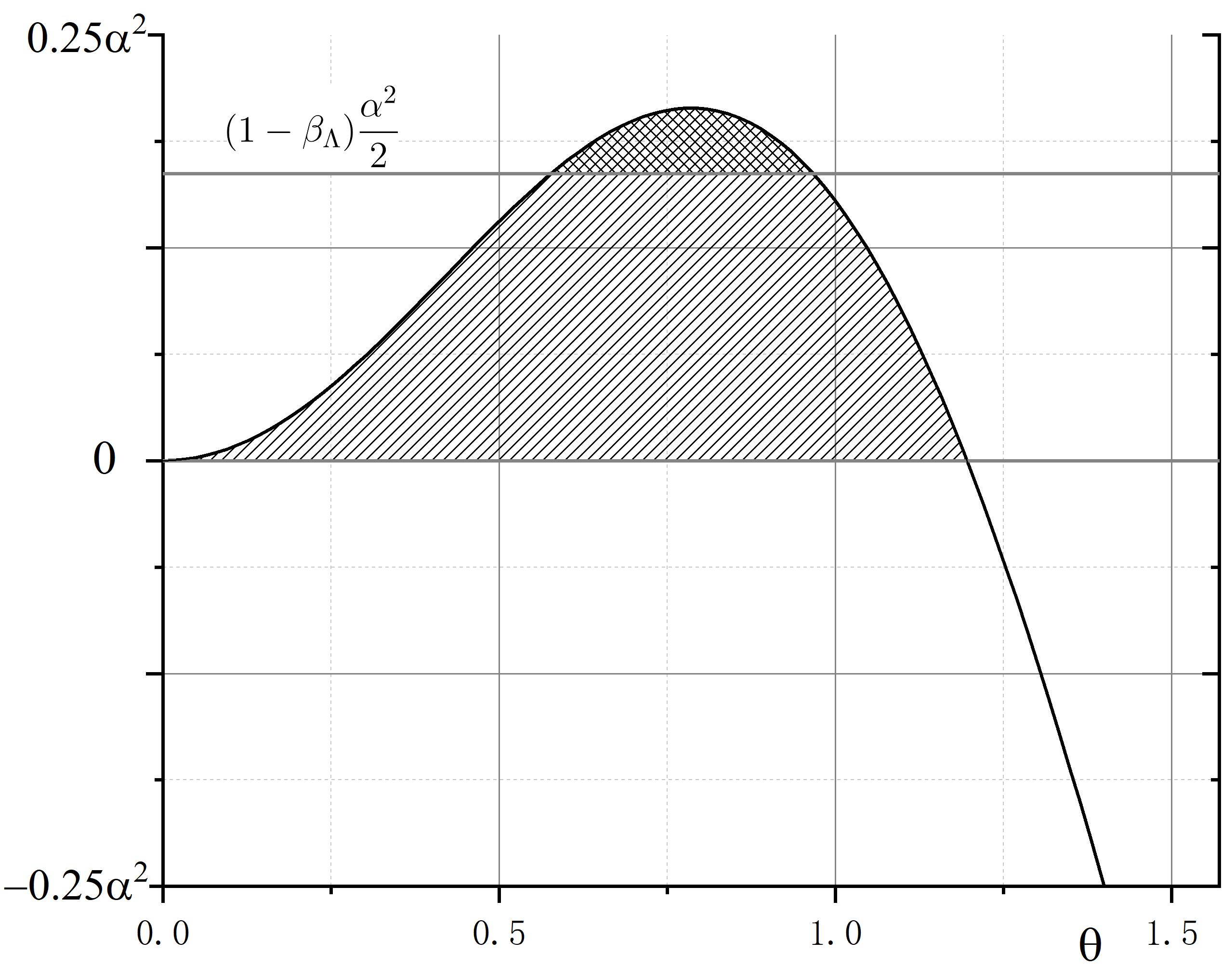}}
\caption{The violation of the generalized CH inequality in the hyperon decay. The generalized CH inequality has an upper bound of 0 which is violated by the quantum mechanics (shaded and doubly shaded region) in a wide range of the parameter $\theta$. For the massive quanta in $\eta_c\to \Lambda\bar{\Lambda}$, the upper bound becomes $(1-\beta_{\Lambda})\alpha^2/2$, which is violated only in the doubly shaded region.}
\label{Figure-CH-V}
\end{figure}

In the process $\eta_c \to \Lambda  \bar{\Lambda}$, the spin state of the hyperon pair is $\rho_{12} = |\psi_{12}\rangle \langle \psi_{12}|$ and
\begin{eqnarray}
|\psi_{12}\rangle =\frac{1}{\sqrt{2}}(|+-\rangle- |-+\rangle) \, .
\end{eqnarray}
For this spin state $\rho_{12}$ and considering the $CP$ conservation decay parameter $\alpha$, the quantum mechanical predictions of Eq. (\ref{P-n1n2-Q}) and (\ref{P-n1-Q}) give
\begin{align}
P(\vec{n}_1,\vec{n}_2) = \frac{1}{4}(1+\alpha^2\vec{n}_1\cdot\vec{n}_2) \; , \; P(\vec{n}_1) = \frac{1}{2}\; , \label{Joint-Dis-QM}
\end{align}
where $P(\vec{n}_1,\vec{n}_2)$ represents the joint probability for the proton $p$ coming along $\vec{n}_1$ and the antiproton $\bar{p}$ coming along $\vec{n}_2$ in the subsequent decay $\Lambda\bar{\Lambda}\to (p\pi^-)(\bar{p}\pi^+)$. A similar expression for the joint decay distribution $\mathcal{I}(\vec{n}_1, \vec{n}_2)$ of the final state momentums has been studied in Ref. \cite{Tornqvist-1981}, i.e.,
\begin{align}
\mathcal{I}(\vec{n}_1,\vec{n}_2) \propto 1 + \alpha^2\vec{n}_1\cdot\vec{n}_2 \; . \label{Decay-dist-ext}
\end{align}
Due to the CHSH inequality for correlation functions \cite{CHSH}, the correlation term shall be extracted from the right-hand side of Eq. (\ref{Decay-dist-ext})  \cite{Tornqvist-1981}. Great progress has been made along the lines of exploring the Bell inequalities involving the correlation functions \cite{Baranov,ptep}.

The CH inequality of Eq. (\ref{CH-lambda-bar}) has the merit that it involves only joint distributions and no further extraction of correlations is needed here. Hence, the contradiction between the local realism and quantum mechanics can be verified directly by taking the quantum prediction of Eq. (\ref{Joint-Dis-QM}) in the CH inequality of Eq. (\ref{CH-lambda-bar}),
\begin{eqnarray}\label{GCH-3}
\frac{\alpha^2}{4}(\cos{\theta_{12}}- \cos{\theta_{12'}}+ \cos{\theta_{1'2}}+\cos{\theta_{1'2'}}) - \frac{\alpha^2}{2} \leq 0\; , \label{CH-alpha-V}
\end{eqnarray}
where $\theta_{ij}$ are the angles between $\vec{n}_i$ and $\vec{n}_j$. Due to the symmetric properties of $P(\vec{n}_1,\vec{n}_2)$, $\theta_{ij}$ may be extended to the range of $[0,2\pi]$ with $\theta_{ij}$ being equivalent to $(2\pi - \theta_{ij})$. For $\theta_{12}=\theta_{1'2}=\theta_{1'2'}= \theta$ and $\theta_{12'}= 3\theta$, we arrive at the following inequality:
\begin{align}
\alpha^2\left[\frac{3\cos\theta-\cos(3\theta)}{4} - \frac{1}{2} \right] \leq 0 \; . \label{CH-theta}
\end{align}
The violation of the inequality (\ref{CH-theta}) is plotted in Fig. \ref{Figure-CH-V}. The maximal violation happens at $\theta =\pi/4$ where
\begin{align}
\alpha^2\left(\frac{\sqrt{2}}{2} - \frac{1}{2}\right) \leq 0\;. \label{Max-value}
\end{align}
Hence, the quantum predictions violate the local realism, and the violation scales with the square of the decay parameter $\alpha$ (see the shaded and doubly shaded region in Fig. \ref{Figure-CH-V}).

\section{Testing the violation in experiments}

\subsection{Measurement of the joint distribution}

From Eq. (\ref{QM-Measurement}), we may rewrite Eq. (\ref{diff-lambda-ppi}) as
\begin{align}
\frac{\mathrm{d} \sigma_{\Lambda \to p\pi^-}}{\mathrm{d}\vec{n}_{p}} = \frac{1}{2\pi} \langle \psi|M_+^{\dag}(\vec{n_p}) M_+(\vec{n}_p)|\psi\rangle \; ,
\end{align}
where $|\psi\rangle$ is the polarization state of $\Lambda$. In the process of $\eta_c\to \Lambda\bar{\Lambda} \to (p\pi^-)(\bar{p}\pi^+)$, the joint distribution may be expressed as
\begin{align}
\frac{\mathrm{d}\sigma_{\eta_c\to\Lambda\bar{\Lambda}\to(p\pi^-)(\bar{p}\pi^+) }}{\mathrm{d}\vec{n}_p \mathrm{d}\vec{n}_{\bar{p}}} = &\frac{1}{4\pi^2} \mathrm{Tr}\left [\rho_{12}(M_+^{(1)\dag}M^{(1)}_+)\otimes (M_+^{(2)\dag}M^{(2)}_+)\right] \nonumber \\
 = & \frac{1}{4\pi^2} P(\vec{n}_p,\vec{n}_{\bar{p}}) \; . \label{joint-diff}
\end{align}
Here $\rho_{12}$ is the polarization state of the bipartite system of $\Lambda\bar{\Lambda}$; $\vec{n}_{p}$ and $\vec{n}_{\bar{p}}$ are the unit directions of the momenta of $p$ and $\bar{p}$ in the rest frame of the $\Lambda$ and $\bar{\Lambda}$ respectively. According to Eq. (\ref{joint-diff}), the joint distribution of $P(\vec{n}_p,\vec{n}_{\bar{p}})$ can be measured from the differential cross section. When putting the joint distributions into Eq. (\ref{CH-lambda-bar}), a violation is expected from the quantum mechanics.

Note the operators $M_{\pm}$ are regarded as general measurements with dichotomic outcomes in the polarized $\Lambda$ decay. Nonetheless, how the measurements are actually implemented during the weak decay $\Lambda\to p\pi^-$ is not relevant in testing the local realism via the generalized CH inequality. Another important requirement in testing the local realism versus the predictions of quantum mechanics is that one cannot refer to the results deduced from quantum theory itself. That is, in the bipartite $\Lambda\bar{\Lambda}$ system, renormalization of the spin correlation function by the asymmetry parameter $\alpha$ is not allowed \cite{Hiesmayr-2015}. It is clear that no such renormalization problem exists in Eq. (\ref{joint-diff}). Finally, we note that the active control problem still remains, and readers may refer to Ref. \cite{YS-JY-L} for a recent discussion.

\subsection{Space-like separation}

In order to test the nonlocal correlation, any possible classical communication should be excluded; i.e., the decays need to be spacelike separated. In the process of $\eta_c\rightarrow \Lambda \bar{\Lambda}$, because $\Lambda$ and $\bar{\Lambda}$ are flying apart at a speed of $v<c$, not all the subsequent decay events $\Lambda\to p \pi^-$ and $\bar{\Lambda}\to \bar{p}\pi^+$ are spacelike separated. Suppose they decay at positions $x_1$ and $x_2$ on each side, respectively; the two events are space-like separated if there is a time interval during which information cannot be communicated at the speed of light $c$,
\begin{align}
c\left|\frac{x_1}{v}-\frac{x_2}{v}\right| \leq x_1+x_2 \; . \label{Space-like-vc}
\end{align}
Here $v$ is the speed of $\Lambda(\bar{\Lambda})$ in the rest frame of $\eta_c$. Eq. (\ref{Space-like-vc}) can be simplified to
\begin{eqnarray}
\frac{1}{k} \leq \frac{x_1}{x_2} \leq k\, ,
\end{eqnarray}
where $k=\frac{1 + \beta_{\Lambda}}{1-\beta_{\Lambda}}$, $\beta_{\Lambda}=v/c$. The fraction of the space-like separated events to the total events of hyperon pairs is as follows:
\begin{eqnarray}
F = \int^\infty_0 e^{-x_2} d x_2 \int^{k x_2}_{\frac{1}{k} x_2} e^{-x_1} d x_1 =\beta_{\Lambda} \;.
\end{eqnarray}
For timelike events (fraction of $1-\beta_{\Lambda}$), the left-hand side of inequality (\ref{CH-lambda-bar}) may reach the maximal value of
\begin{align}
\frac{1+\alpha^2}{4} -\frac{1-\alpha^2}{4} + \frac{1+\alpha^2}{4} + \frac{1+\alpha^2}{4} -\frac{1}{2} - \frac{1}{2} + \frac{1-\alpha^2}{2} = \frac{\alpha^2}{2}  \;.
\end{align}
Therefore the realism prediction of Eq. (\ref{CH-lambda-bar}) now turns to
\begin{align}
P(\vec{n}_1,\vec{n}_2)-P(\vec{n}_1,\vec{n'}_2)+P(\vec{n'}_1,\vec{n}_2)+P(\vec{n'}_1,\vec{n'}_2)& \nonumber \\
-P(\vec{n'}_1)-P(\vec{n}_2)+\frac{1 - \alpha^2}{2} & \leq
\beta_{\Lambda} \cdot 0+(1-\beta_{\Lambda}) \frac{\alpha^2}{2} \; . \label{CH-alpha-beta}
\end{align}
From Eq. (\ref{Max-value}), we find that the contradiction between the quantum prediction and the local realism is still observable in the case
\begin{align}
\beta_{\Lambda} \cdot 0+(1-\beta_{\Lambda}) \frac{\alpha^2}{2} < \alpha^2 \frac{\sqrt{2}-1}{2} \;, \label{P-R-condition}
\end{align}
which gives $\beta_{\Lambda} >2-\sqrt{2}\sim 0.586$, while the ratio of spacelike $\Lambda\bar{\Lambda}$ from $\eta_c$ is $\beta_{\Lambda}=0.664$ \cite{pdg}. The violation of Eq. (\ref{CH-alpha-beta}) is also presented in Fig. \ref{Figure-CH-V}, where the upper bound changes from 0 to $(1-\beta_{\Lambda})\displaystyle \frac{\alpha^2}{2}$.

\section{Discussions}

In this work, we present a generalized CH inequality for the measurements whose outcome probabilities do not span the whole range of $[0,1]$. Since the differential decay amplitudes in high energy physics usually have a similar behavior to that of $\Lambda\to p\pi^-$, which we are concerned with in this work, the generalized CH inequality provides a proper formalism to compare the local realism and quantum predictions, and is applicable to a wide range of interactions. The massive entangled quanta experience both weak and strong interactions here; hence, the quantum nonlocality in the $\eta_c \to \Lambda\bar{\Lambda}$ induced process, per se, is typically important.

By embedding the polarization distributions of entangled $\Lambda \bar{\Lambda}$ pairs into the generalized CH inequality and taking typical correlation angles, it is found that the quantum theory calculation leads to an evident violation of the generalized CH inequality. This can be readily examined in experiment, like at BES~III where a collection of $10^8$ $\eta_c$ events was obtained \cite{FPPBESSIII}. Given the branching ratios in the decay chain, namely, $Br(\eta_c \to \Lambda \bar{\Lambda})\approx (1.09\pm 0.24)\times 10^{-3}$ and $Br(\Lambda \to p \pi^-) \approx (63.9\pm 0.5)\%$ \cite{pdg}, there will be more than tens of thousands events that can be employed to check the inequality. Even by taking account of the 10\% detection efficiency~\cite{ronggang}, one can still perform such an experiment. It is worth emphasizing that the novel CH inequality has the merit of not being subject to the result deduced from quantum theory calculations like Bell or CHSH inequalities, which are about correlation functions.

Finally, we note that although the high energy physics experiments are generally lacking free will in testing quantum correlation, which impairs the steerability to refute the local realism, the quantum Bell nonlocality can still be examined.

\section*{Acknowledgements}
\noindent
We thank Qun Wang, Rong-Gang Ping, and Yang-Guang Yang for discussions. This work was supported in part by the National Natural Science Foundation of China(NSFC) under Grants No. 11975236, No. 11635009, No. 11375200, No. 11535012, and No. 11890713.

\end{document}